\documentclass{article}
\usepackage{spconf,amsmath,graphicx}
\usepackage{subcaption}
\usepackage{float}
\usepackage[normalem]{ulem}
\usepackage[utf8]{inputenc}
\usepackage{hyperref}

\useunder{\uline}{\ul}{}

\title{ESRGAN+ : Further Improving Enhanced Super-Resolution Generative Adversarial Network}
\name{Nathanaël Carraz Rakotonirina, Andry Rasoanaivo}

\address{Laboratoire d'Informatique et Mathématiques, Université d'Antananarivo, Madagascar}
\begin{document}
%\ninept
%
\maketitle
\begin{abstract}
	Enhanced Super-Resolution Generative Adversarial Network (ESRGAN) is a perceptual-driven approach for single image super-resolution that is able to produce photorealistic images. Despite the visual quality of these generated images, there is still room for improvement. In this fashion, the model is extended to further improve the perceptual quality of the images. We have designed a network architecture with a novel basic block to replace the one used by the original ESRGAN. Moreover, we introduce noise inputs to the generator network in order to exploit stochastic variation. The resulting images present more realistic textures. The code is available at \url{https://github.com/ncarraz/ESRGANplus}.
\end{abstract}
\section{Introduction}
\label{sec:intro}

Super-resolution (SR) is the task of generating a high-resolution (HR) image using low-resolution (LR) ones. When only one LR image is used, it is commonly called Single Image Super-Resolution (SISR). The target of such task used to be the minimization of the mean squared error (MSE) between the generated image and the original one. This results in maximizing the peak signal-to-ratio (PSNR) which is a standard measure for SISR. However, PSNR-oriented approaches do not generate perceptually good images \cite{ledig2017photo}. Perceptual-oriented methods were then proposed. Super-Resolution Generative Adversarial Network (SRGAN) \cite{ledig2017photo} uses both perceptual loss \cite{johnson2016perceptual,bruna2015super} and generative adversarial networks (GANs) \cite{goodfellow2014generative} to produce images residing in the manifold of natural images. Enhanced Super-Resolution Generative Adversarial Network (ESRGAN) \cite{wang2018esrgan} improves SRGAN by introducing an architecture composed of Residual-in-Residual Dense
Blocks (RRDB) without Batch Normalization (BN) \cite{ioffe2015batch} layers. Besides, relativistic average GAN (RaGAN) \cite{jolicoeur2018relativistic} was used as the discriminator and the features were used before activation.

We aim to further improve the perceptual quality of the images generated by ESRGAN. First, we propose a new block called Residual-in-Residual Dense Residual Block (RRDRB) which has higher capacity than ESRGAN's RRDB block. Second, we introduce noise inputs in the network as in \cite{karras2018style} in order to benefit from stochastic variation.

\section{Related work}
\label{sec:related_work}

The main approaches to SISR can be divided into three distinct categories: interpolation-based
methods, reconstruction-based methods and learning-based methods \cite{yang2018deep}. Approaches based on deep learning have further surpassed the two former methods as well as simple learning-based methods.

The very first deep learning-based approach, proposed by Dong et al. \cite{dong2014learning,dong2016image}, is SRCNN. It makes use of convolutional neural networks in an end-to-end manner. Though the network is shallow, it outsmarted previous techniques as far as the SISR task is concerned. Kim et al. \cite{kim2016accurate} introduce a deeper model called VDSR. With a similar performance, DRCN \cite{kim2016deeply} exploits deep recursive networks by combining intermediary results. SRResNet \cite{ledig2017photo} and DRRN \cite{tai2017image} make use of residual units. EDSR \cite{huang2017densely} along with MDSR, its multiple scale factors version, are the state-of-the-art methods for PSNR-based super-resolution. Residual dense networks were used in SRDenseNet and Memnet \cite{tong2017image}. 

In order to focus more on the visual quality of generated images, a perceptual loss closer to perceptual quality is proposed. SRGAN which is based on GANs  uses this perceptual loss along with adversarial loss to produce photo-realistic images. These images are visually more convincing despite having lower score on standard quantitative measure like PSNR and structural similarity (SSIM). EnhanceNet \cite{sajjadi2017enhancenet} is also based on GANs but uses a different architecture. ESRGAN as its name implies enhances SRGAN. It introduced a new block with a higher capacity named RRDB. Besides BN layers were removed, residual scaling \cite{szegedy2017inception} and smaller initialization were used to facilitate training a very deep network. The discriminator uses relativistic average GAN, which
learns to evalate ``whether one image is more realistic than the other'' rather than
``whether one image is real or fake''. Furthermore, in the perceptual loss, the VGG features are taken before activation rather than after as in SRGAN. There is still a gap between ground-truth images and images generated by ESRGAN. The present work aims to further close this gap.

\section{Method}
\label{sec:method}

\subsection{Network architecture}
\label{ssec:network_arch}
\begin{figure}[htb]
	\centering
	\includegraphics[width=8.5cm]{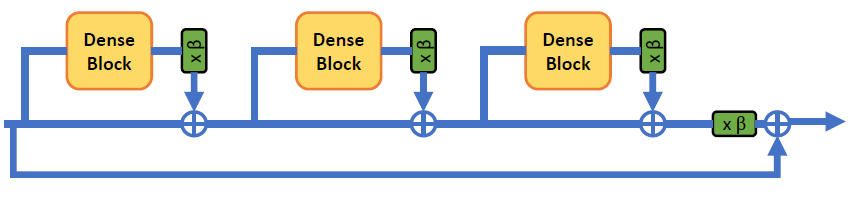}
	\caption{The basic block used in ESRGAN called Residual in Residual Dense Block (RRDB).}
	\label{fig:rrdb}
\end{figure}

ESRGAN's basic block allows the network to be easier to train and have a very high capacity. The overall architecture of ESRGAN is maintained as depicted in Figure \ref{fig:rrdb} except for the Dense block which is replaced by our new block. 

\begin{figure}[htb]
	\centering
	\begin{subfigure}{.5\textwidth}
		\centering
		\includegraphics[width=8.5cm]{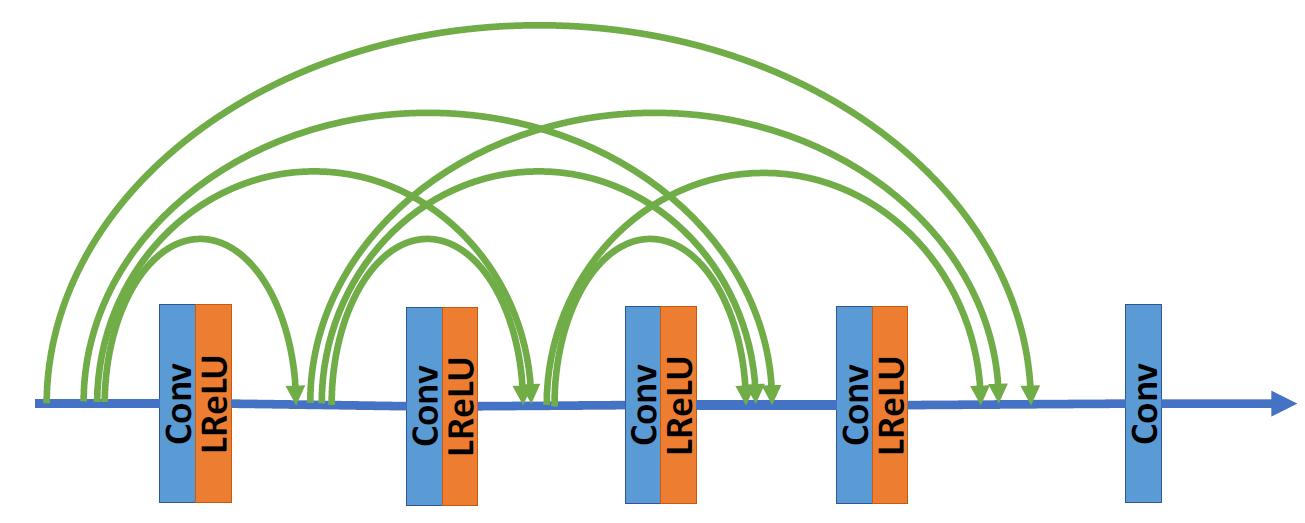}
		\caption{Dense block}
		\label{fig:sub1}
	\end{subfigure}
	\begin{subfigure}{.5\textwidth}
		\centering
		\includegraphics[width=8.5cm]{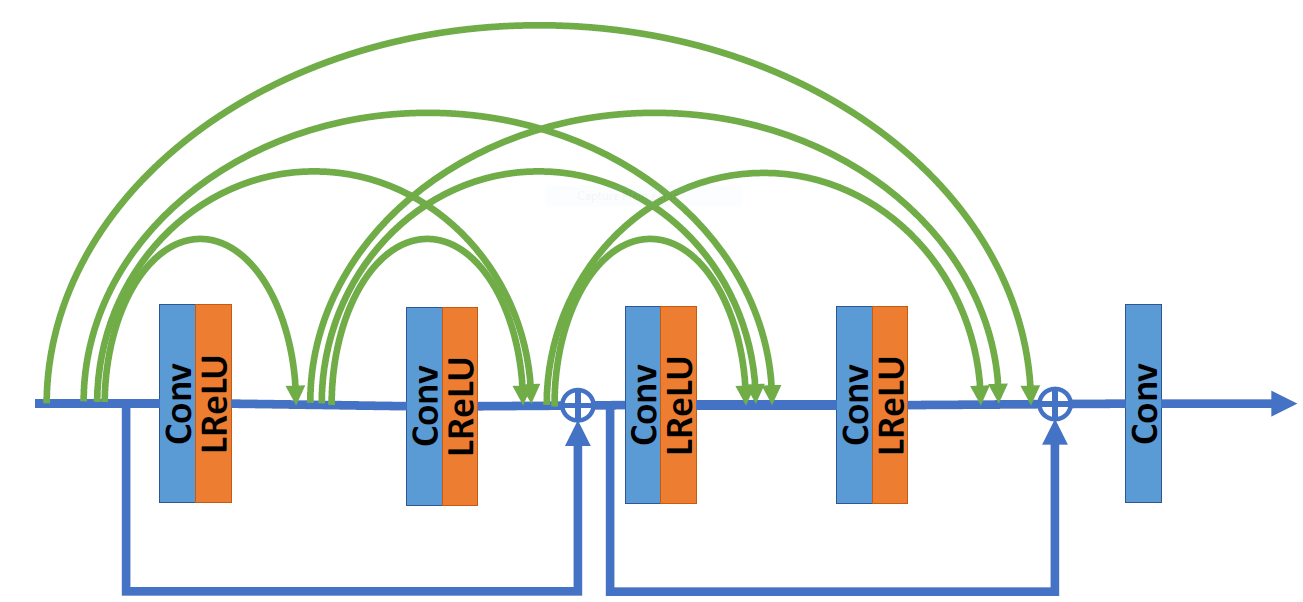}
		\caption{Residual Dense block}
		\label{fig:sub2}
	\end{subfigure}
	\caption{\textbf{Top:} Dense block is the main path used in ESRGAN's RRDB. \textbf{Bottom:} Residuals are added every two layer in the Dense block.}
	\label{fig:block_comparison}
\end{figure}

The novel block we propose results in greater capacity. RRDB has a residual-in-residual structure with Dense blocks \cite{huang2017densely} in the main path. We add an additional level of residual learning inside the Dense blocks as presented in Figure \ref{fig:block_comparison} to augment the network capacity without increasing its complexity. A residual is then added every two layers in each Dense block. The visual quality of the generated images using the new block is substantially superior to that of the simple Dense block. As described in \cite{chen2017dual}, ResNet enables to re-use features while DenseNet enables to find new features. This new architecture then benefits from both feature exploitation and exploration resulting in images of superior perceptual quality. We name ESRGAN+ the model using this new architecture.

\subsection{Noise inputs}
\label{ssec:noise_inputs}

\begin{figure}[htb]
	\centering
	\includegraphics[width=8.5cm]{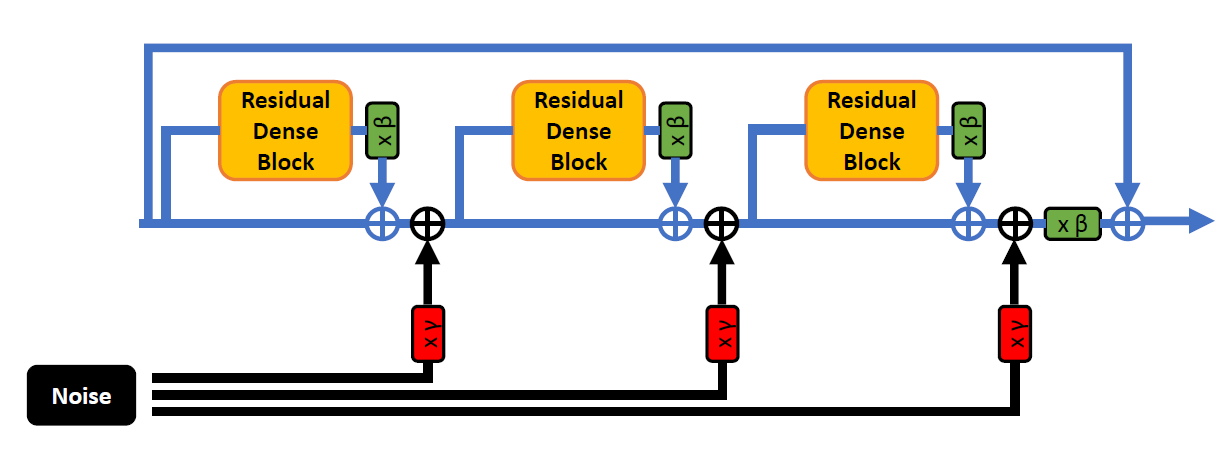}
	\caption{Gaussian noise is added after each residual along with a learned scaling-factor.}
	\label{fig:noise_per_residual_dense_block}
\end{figure}

Adding noise to the generator was recently used in human faces generation \cite{karras2018style} which also heavily relies on GANs. However, it was never applied to super-resolution. In order to have stochastic detail, noise inputs are introduced in the generator's architecture. Gaussian noise is added to the output of each residual dense block along with learned per-feature scaling factors , as illustrated in Figure \ref{fig:noise_per_residual_dense_block}. %We expose the results of using different configurations with the noise inputs in \ref{appendix:noise}.

Stochastic variation randomizes only certain local aspects of the generated images without changing our global perception of the images \cite{karras2018style}. The effects of the noise inputs are very localized leaving intact the general structure and the higher level information of the images. The network does not need to generate spatially-varying pseudorandom numbers when that is required. Consequently, the network capacity that would have been wasted for that task can be efficiently used to give finer-details in the high-level aspects. The model using both the new block and the noise inputs is called nESRGAN+.

\begin{table*}[t]
	\caption{Quantitative evaluation of our models with other perceptual-driven methods. The best and second best results are highlighted and underlined, respectively. We evaluate using the perceptual index (value on the right) but PSNR (value on the left) is also given for reference purposes.}
	\label{table:comparison_pirm}
	
	\centering
	\begin{tabular}{p{3cm}|p{3cm}|p{3cm}|p{3cm}|p{3cm}|}
		\cline{2-5}
		& EnhanceNet          & ESRGAN     & ESRGAN+ (ours)      & nESRGAN+ (ours)     \\ \hline
		\multicolumn{1}{|l|}{Validation PIRM} & 25.06/2.68          & 25.17/2.55 & 24/{\ul2.38}       & 24.32/\textbf{2.36} \\
		\multicolumn{1}{|l|}{Test PIRM}       & 24.94/2.72          & 25.03/2.43 & 23.80/{\ul2.31}    & 24.15/\textbf{2.29} \\
		\multicolumn{1}{|l|}{Urban100}        & 23.54/\textbf{3.47} & 24.36/3.77 & 23.28/{\ul3.55}    & 23.22/3.55          \\
		\multicolumn{1}{|l|}{OST300}          & 24.37/2.82          & 24.64/2.49 & 23.84/\textbf{2.46} & 23.80/{\ul 2.49}    \\ \hline
	\end{tabular}
	
\end{table*} 

\captionsetup[subfigure]{labelformat=empty}
\begin{figure*}
	\begin{subfigure}{\textwidth}
		\begin{subfigure}{.25\textwidth}
			\centering
			\includegraphics[width=\textwidth]{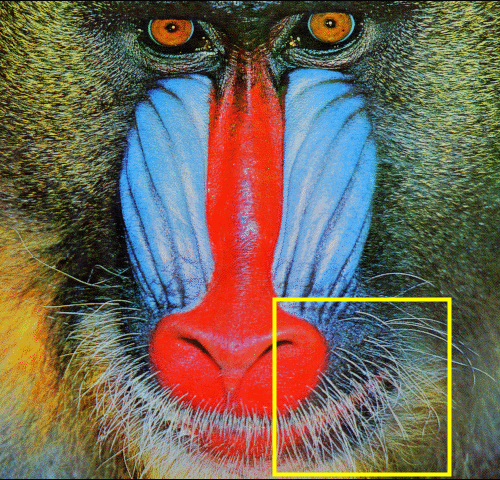}
			\caption{baboon from Set14}
		\end{subfigure}
		~~
		\begin{subfigure}{.75\textwidth}
			\begin{subfigure}{.23\textwidth}
				\centering
				\includegraphics[width=\textwidth]{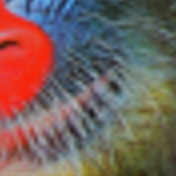}
				\caption{Bicubic\\ (22.43/6.77)}
			\end{subfigure}
			~
			\begin{subfigure}{.23\textwidth}
				\centering
				\includegraphics[width=\textwidth]{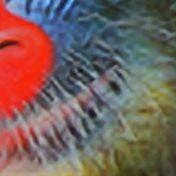}
				\caption{SRCNN \\ (22.70/5.89)}
			\end{subfigure}
			~
			\begin{subfigure}{.23\textwidth}
				\centering
				\includegraphics[width=\textwidth]{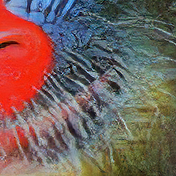}
				\caption{EnhanceNet \\ (20.87/2.65)}
			\end{subfigure}
			~
			\begin{subfigure}{.23\textwidth}
				\centering
				\includegraphics[width=\textwidth]{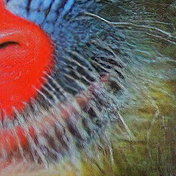}
				\caption{SRGAN \\ (21.14/2.61)}
			\end{subfigure}
			
			\begin{subfigure}{.23\textwidth}
				\centering
				\includegraphics[width=\textwidth]{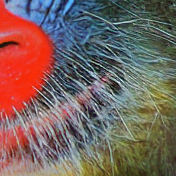}
				\caption{ESRGAN \\ (20.32/1.99)}
			\end{subfigure}
			~
			\begin{subfigure}{.23\textwidth}
				\centering
				\includegraphics[width=\textwidth]{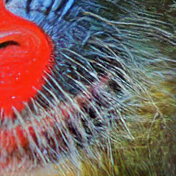}
				\caption{ESRGAN+ \\ (19.79/1.81)}
			\end{subfigure}
			~
			\begin{subfigure}{.23\textwidth}
				\centering
				\includegraphics[width=\textwidth]{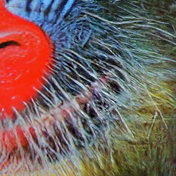}
				\caption{nESRGAN+ \\ (19.71/2.14)}
			\end{subfigure}
			~
			\begin{subfigure}{.23\textwidth}
				\centering
				\includegraphics[width=\textwidth]{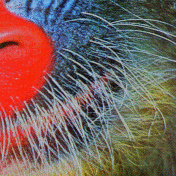}
				\caption{HR \\ ($\infty$/3.59)}
			\end{subfigure}
		\end{subfigure}
	\end{subfigure}
	~
	
	\begin{subfigure}{\textwidth}
		\begin{subfigure}{.25\textwidth}
			\centering
			\includegraphics[width=\textwidth]{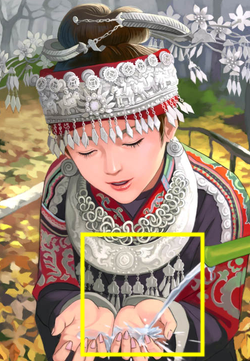}
			\caption{comic from Set14}
		\end{subfigure}
		~~
		\begin{subfigure}{.75\textwidth}
			\begin{subfigure}{.23\textwidth}
				\centering
				\includegraphics[width=\textwidth]{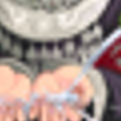}
				\caption{Bicubic \\ (21/6.73)}
			\end{subfigure}
			~
			\begin{subfigure}{.23\textwidth}
				\centering
				\includegraphics[width=\textwidth]{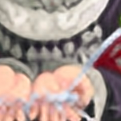}
				\caption{SRCNN \\ (22.52/6.74)}
			\end{subfigure}
			~
			\begin{subfigure}{.23\textwidth}
				\centering
				\includegraphics[width=\textwidth]{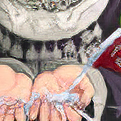}
				\caption{EnhanceNet \\ (20.64/2.40)}
			\end{subfigure}
			~
			\begin{subfigure}{.23\textwidth}
				\centering
				\includegraphics[width=\textwidth]{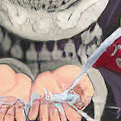}
				\caption{SRGAN \\ (19.32/2.32)}
			\end{subfigure}
			
			\begin{subfigure}{.23\textwidth}
				\centering
				\includegraphics[width=\textwidth]{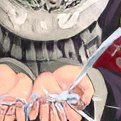}
				\caption{ESRGAN \\ (18.87/2.27)}
			\end{subfigure}
			~
			\begin{subfigure}{.23\textwidth}
				\centering
				\includegraphics[width=\textwidth]{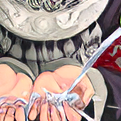}
				\caption{ESRGAN+ \\ (18.06/2.17)}
			\end{subfigure}
			~
			\begin{subfigure}{.23\textwidth}
				\centering
				\includegraphics[width=\textwidth]{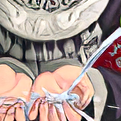}
				\caption{nESRGAN+ \\ (17.76/2.63)}
			\end{subfigure}
			~
			\begin{subfigure}{.23\textwidth}
				\centering
				\includegraphics[width=\textwidth]{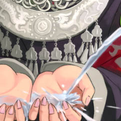}
				\caption{HR \\ ($\infty$/2.76)}
			\end{subfigure}
		\end{subfigure}
	\end{subfigure}
	
	\caption{Comparison between the qualitative results of the main perceptual-driven models and ESRGAN+ using images from Set14. PNSR (value on the left) and perceptual index (value on the right) are used for the evaluation.}
	\label{fig:qualitative_results_baboon}
\end{figure*}

\section{Experiments}
\label{sec:experiments}

\subsection{Data}
\label{ssec:data}

The used training set is DIV2K \cite{agustsson2017ntire}. It is a dataset of 2K resolution images adequate for the task of SR. Originally, there are only 800 images in the DIV2K dataset. As in ESRGAN, data augmentation is performed through random horizontal 
flips and rotations. The benchmark datasets used for evaluation are BSD100 \cite{martin2001database}, Urban100 \cite{huang2015single}, OST300 \cite{wang2018sftgan}, Set5 \cite{bevilacqua2012low}, Set14 \cite{zeyde2010single} and the PIRM datasets \cite{blau20182018}.

\subsection{Training details and parameters}
\label{ssec:training}

The LR images are obtained
by downsampling the HR images using
bicubic kernel with a scaling factor of x4. We maintain all the training parameters of the original ESRGAN. We crop 128 x 128 HR sub images. The size of the mini-batch is 16. A PSNR-oriented pre-trained model is used to initialize the generator. The loss function remains unchanged with $\lambda = 5 \times 10^{-3}$ and $\eta=1 \times10^{-2}$. The learning rate is set to $1 \times 10^{-4}$ and halved at [50k,100k,200k,300k] iterations.

The model is optimized using Adam with $\beta_1 = 0.9$ and $\beta_2=0.999$. The trained model is the one with the 23 blocks generator. The implementation is done with Pytorch and the training with NVIDIA Tesla K80 GPUs.

\subsection{Results}
\label{sec:results}

We evaluate our two models with other perceptual-driven approaches on the PIRM datasets (see Table \ref{table:comparison_pirm}). %Evaluation on the benchmak datasets of other models including those based on PSNR is presented in \ref{appendix:comparison}. 
In the YCbCr color space, PSNR is measured on the luminance channel. The perceptual index is the one used in the PIRM-SR Challenge \cite{blau20182018}. It is based on the Ma's score \cite{ma2017learning} and NIQE \cite{mittal2012making} and equals $\frac{1}{2}((10-Ma)+NIQE)$. Higher is better when measuring with the PSNR whereas lower is better when considering the perceptual index. Both of our models always perform better compared to ESRGAN. We see that nESRGAN+ has a better perceptual score on the PIRM datasets. This highlights the benefits of using the noise inputs in the generator network. However, there are still limitations associated with the noise injection's generalization. Adding noise does not always result in better perceptual quality. This is the case for categories of images which do not fully exploit stochastic variation such as images of buildings in Urban100 and OST300. Future works will focus on getting the most out of the Gaussian noise.

Qualitative comparison is made in Figure \ref{fig:qualitative_results_baboon} between our models and others based on PSNR and perceptual quality such as SRCNN, EnhanceNet, SRGAN, ESRGAN using images from the Set14 dataset. It can be observed that the images reconstructed by our models present more detailed structures and are less distinguishable from the ground truth images when compared to the other pictures. Most of the original textures are kept like the boy's complexion. %We provide additional visual results in \ref{appendix:visual_results}. 

\section{Conclusion}
\label{sec:conclusion}

We have proposed ESRGAN+ and nESRGAN+ which outperform other approaches as long as perceptual quality is concerned. A new basic block has been introduced to further increase the capacity of the network. Moreover, noise inputs are added to benefit from stochastic variation. All these improvements have contributed to the generation of images with more natural textures as well as greater sharpness and details.

\bibliographystyle{IEEEbib}
\bibliography{lib}

\end{document}